\documentclass[twocolumn,english,aps,prl,reprint,groupedaddress,notitlepage,nobibnote,nofootinbib,preprintnumbers,showpacs]{revtex4-1}
\pdfoutput=1
\usepackage{lmodern}

\usepackage[T1]{fontenc}
\usepackage[latin9]{inputenc}
\usepackage{geometry}
\usepackage{color}
\usepackage{babel}
\usepackage{amsmath}
\usepackage{amssymb}
\usepackage{graphicx}
\usepackage{esint}
\usepackage{wasysym}
\usepackage{array}
\usepackage{comment}
\usepackage{tabu}
\usepackage{relsize}

\usepackage[table]{xcolor}
\usepackage[unicode=true,pdfusetitle,
 bookmarks=true,bookmarksnumbered=false,bookmarksopen=false,
 breaklinks=false,pdfborder={0 0 1},backref=false,colorlinks=true]
 {hyperref}

\hypersetup{
 citecolor=blue,linkcolor=blue,urlcolor=blue}
\makeatletter

 \@ifundefined{textcolor}{}
 {
   \definecolor{BLACK}{gray}{0}
   \definecolor{WHITE}{gray}{1}
   \definecolor{RED}{rgb}{1,0,0}
   \definecolor{GREEN}{rgb}{0,1,0}
   \definecolor{BLUE}{rgb}{0,0,1}
   \definecolor{CYAN}{cmyk}{1,0,0,0}
   \definecolor{MAGENTA}{cmyk}{0,1,0,0}
   \definecolor{YELLOW}{cmyk}{0,0,1,0}
 }
\usepackage{babel}

\newcommand{\be}{\begin{equation}}
\newcommand{\ee}{\end{equation}}
\newcommand{\bea}{\begin{eqnarray}}
\newcommand{\eea}{\end{eqnarray}}

\newcommand{\Fig}[1]{Fig.~\ref{#1}}
\newcommand{\Eq}[1]{Eq.~(\ref{#1})}
\newcommand{\Eqs}[2]{Eqs.~(\ref{#1}) and (\ref{#2})}

\newcommand{\nda}{\scriptstyle{\times}}
\newcommand{\ysq}{\scriptstyle{y^2}}
\newcommand{\ydsq}{\scriptstyle{\bar y^{2}}}

\newcommand{\cw}{\cellcolor{white}}

\newcommand{\czb}{\cellcolor{zero2}}

\definecolor{zero1}{rgb}{0.35,0.4,0.85}
\definecolor{zero2}{rgb}{0.88,0.88,.88}
\definecolor{zero3}{rgb}{0.85,0.85,0.95}
\definecolor{zero4}{rgb}{1,0.3,0.3}

\newcommand{\nn}{\nonumber}

\makeatletter
\def\CT@@do@color{%
	\global\let\CT@do@color\relax
		\@tempdima\wd\z@
		\advance\@tempdima\@tempdimb
		\advance\@tempdima\@tempdimc
		\advance\@tempdimb\tabcolsep
		\advance\@tempdimc\tabcolsep
		\advance\@tempdima1.5\tabcolsep
	\kern-1.5\@tempdimb
	\leaders\vrule
	\hskip\@tempdima\@plus  1fill
	\kern-1.5\@tempdimc
	\hskip-\wd\z@ \@plus -1fill }
\makeatother

\begin{document}

\preprint{\hbox{CALT-2015-024} }

\title{Non-renormalization Theorems without Supersymmetry}

\author{Clifford Cheung}
\author{Chia-Hsien Shen}
\affiliation{Walter Burke Institute for Theoretical Physics \\California Institute of Technology, Pasadena, CA 91125}
\date{\today}
\email{clifford.cheung@caltech.edu,chshen@caltech.edu}
\begin{abstract}
We derive a new class of one-loop non-renormalization theorems that strongly constrain the running of higher dimension operators in a general four-dimensional quantum field theory.  Our logic follows from unitarity: cuts of one-loop amplitudes are products of tree amplitudes, so if the latter vanish then so too will the associated divergences.   Finiteness is then ensured by simple selection rules that zero out tree amplitudes for certain helicity configurations.
For each operator we  define holomorphic and anti-holomorphic weights, $(w,\overline w) =(n - h,n+h)$, where $n$ and $h$ are the number and sum over helicities of the particles created by that operator.   We argue that an operator ${\cal O}_i$ can only be renormalized by an operator ${\cal O}_j$ if $w_i \geq w_j$ and $\overline w_i \geq \overline w_j$, absent non-holomorphic Yukawa couplings.  These results explain and generalize the surprising cancellations discovered in the renormalization of dimension six operators in the standard model. Since our claims rely on unitarity and helicity rather than an explicit symmetry, they apply  quite generally.

\end{abstract}

\maketitle

\section{Introduction\label{intro}}

Technical naturalness dictates  that all operators not forbidden by symmetry are compulsory---and thus generated by renormalization.  Softened ultraviolet divergences
are in turn a telltale sign of underlying symmetry.
This is famously true in supersymmetry, where holomorphy enforces powerful non-renormalization theorems.

In this letter we derive a new class of non-renormalization theorems for non-supersymmetric theories. Our results apply to the one-loop running of the leading irrelevant deformations of a four-dimensional quantum field theory of marginal interactions,
\bea
\Delta {\cal L} = \sum_{i} c_i {\cal O}_i,
\eea
where ${\cal O}_i$ are higher dimension operators.   At leading order in $c_i$, renormalization induces operator mixing via
\bea
(4\pi)^2 \frac{d c_i}{d\log \mu}  =  \sum_{j} \gamma_{ij} c_j,
\label{eq:gammaij}
\eea
where by dimensional analysis the  anomalous dimension matrix $\gamma_{ij}$ is a function of marginal couplings alone.

The logic of our approach is simple and makes no reference to symmetry.  Renormalization is induced by log divergent amplitudes, which by unitarity have kinematic cuts equal to products of on-shell tree amplitudes~\cite{Bern:1994cg,*Bern:1994zx}. If any of these tree amplitudes vanish, then so too will the divergence.  
Crucially, many tree amplitudes are zero due to helicity selection rules, which {\it e.g.}~forbid the all minus helicity gluon amplitude in Yang-Mills theory.

\begin{figure}[t]

\begin{center}
\includegraphics[width=.5\textwidth]{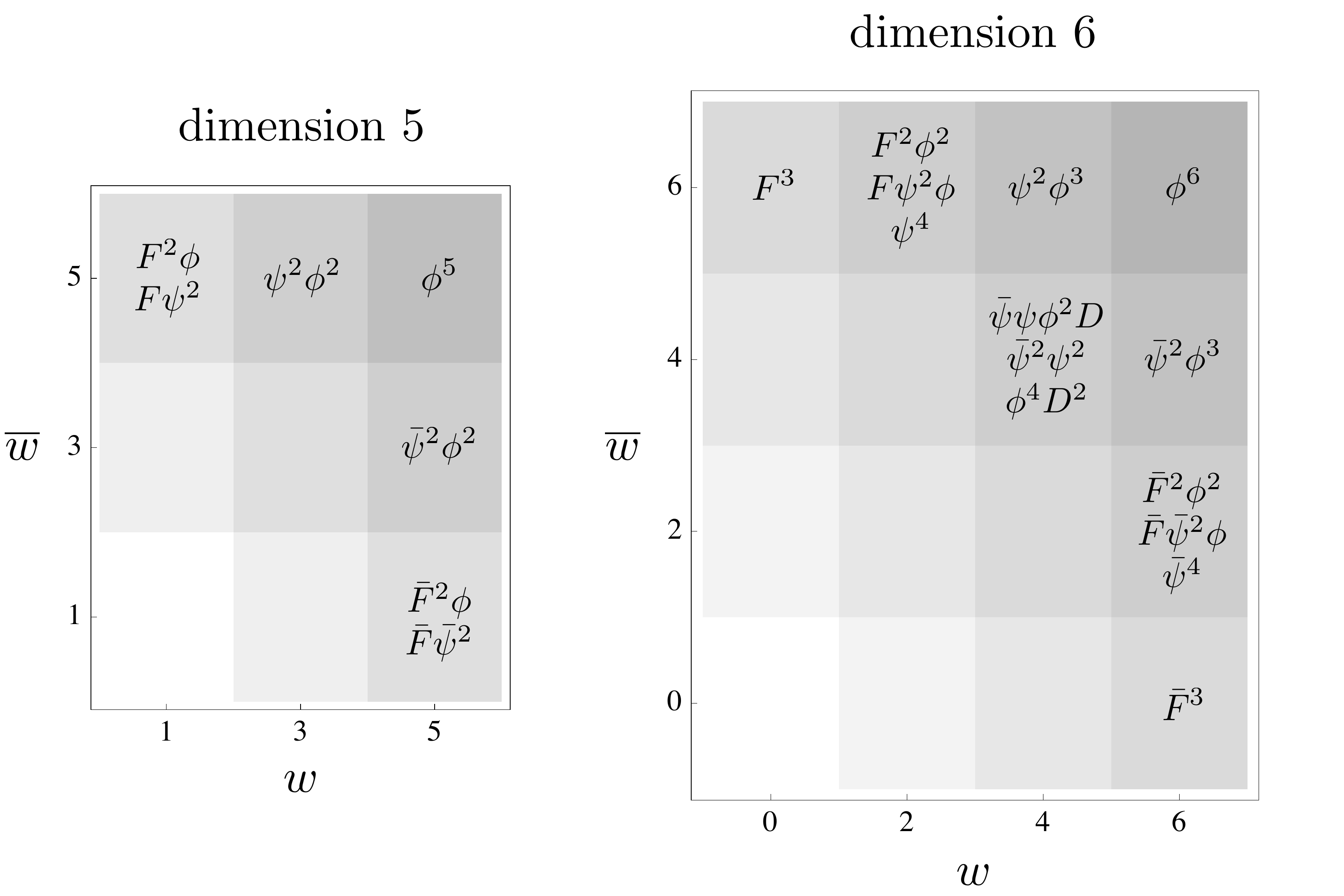}
\end{center}
\vspace*{-.55cm}
\caption{Weight lattice for dimension five and six operators, suppressing flavor and Lorentz structures, {\it e.g.} on which fields derivatives act. The non-renormalization theorems in \Eq{eq:nonrenorm} let operators to mix into operators of equal or greater weight.  Pictorially, this forbids transitions down or to the left.}
\label{fig:wlattice} 
\end{figure}

For our analysis, we define the holomorphic and anti-holomorphic weight of 
an on-shell amplitude $A$ by\footnote{Holomorphic weight is a generalization of $k$-charge in super Yang-Mills theory, where the N$^k$MHV amplitude has $w=k+4$. }
\bea
w(A)= n(A)-h(A), &\qquad& \overline w(A)= n(A)+h(A),
\label{eq:ampweight}
\eea
where $n(A)$ and $h(A)$ are the number and sum over helicities of the external states.    Since $A$ is physical, its weight is field reparameterization and gauge independent.  The weights of an operator ${\cal O}$ are then invariantly defined by minimizing over all amplitudes involving that operator: $w({\cal O}) = \textrm{min}\{ w(A) \}$ and $
\overline w({\cal O}) = \textrm{min}\{ \overline w(A) \}$.
In practice, operator weights are fixed by the leading non-zero contact amplitude\footnote{By definition, all covariant derivatives $D$ are treated as partial derivatives $\partial$ when computing the leading contact amplitude.} built from an  insertion of ${\cal O}$,
\bea
w({\cal O}) =n({\cal O}) -h({\cal O}), &\qquad & \overline w({\cal O}) =n({\cal O}) +h({\cal O}),
\eea
where $n({\cal O})$ is the number of particles created by ${\cal O}$ and $h({\cal O})$ is their total helicity. For field operators we find:
\bea
\setlength{\tabcolsep}{4.5pt}
\renewcommand{\arraystretch}{1.25}
\begin{tabular}{c|c|c|c|c|c|} 
$ {\cal O}$ & $F_{\alpha\beta}$ & $\psi_\alpha$  & $\phi$ & $\bar\psi_{\dot \alpha}$ & $\bar F_{\dot\alpha\dot\beta}$ \\ \hline
$h$ & +1 & $+1/2$  & 0 & $-1/2$ & $-1$ \\ \hline
$(w,\overline w)$ & $(0,2)$ & $(1/2,3/2)$  & $(1,1)$ & $(3/2,1/2)$ & $(2,0)$ \\ \hline
\end{tabular} \nn
\setlength{\tabcolsep}{2pt}
\renewcommand{\arraystretch}{1}
\eea
where all Lorentz covariance is expressed in terms of four-dimensional spinor indices, so {\it e.g.}~the gauge field strength is $F_{\alpha \dot \alpha \beta \dot\beta}= F_{\alpha \beta} \bar\epsilon_{\dot\alpha \dot\beta}+\bar F_{\dot \alpha \dot \beta} \epsilon_{\alpha \beta}$.  The weights of all dimension five and six operators are shown in \Fig{fig:wlattice}.

 
As we will prove, an operator ${\cal O}_i$ can only be renormalized by an operator ${\cal O}_j$ at one-loop if the corresponding weights $(w_i, \overline w_i)$ 
and $(w_j, \overline w_j)$ 
satisfy the inequalities
\bea
w_i \geq w_j \quad &\textrm{and} & \quad \overline w_i \geq \overline w_j,
\label{eq:nonrenorm}
\eea
and all Yukawa couplings are of a ``holomorphic'' form consistent with a superpotential.  This implies a new class of non-renormalization theorems,
\bea
\gamma_{ij} =0 \quad &\textrm{if}& \quad w_i < w_j \quad \textrm{or}  \quad \overline w_i < \overline w_j ,
\label{eq:gammaijzero}
\eea
which dictate mostly zero entries in the anomalous dimension matrix.  The resulting non-renormalization theorems for all dimension five and six operators are shown in Tab.~\ref{tab:dim5} and Tab.~\ref{tab:dim6}.

Since our analysis hinges on unitarity and helicity, the resulting non-renormalization theorems are general and not derived from an explicit symmetry of the off-shell Lagrangian.   Moreover, our findings explain  the ubiquitous and surprising cancellations \cite{Alonso:2014rga} observed in the one-loop renormalization of dimension six operators in the standard model~\cite{Grojean:2013kd,Jenkins:2013zja,*Jenkins:2013wua,*Alonso:2013hga,Elias-Miro:2013gya,*Elias-Miro:2013mua,Elias-Miro:2013eta}.  Lacking an explanation from power counting or spurions, the authors of \cite{Alonso:2014rga} conjectured a hidden ``holomorphy'' enforcing non-renormalization theorems among holomorphic and anti-holomorphic operators.  We show here that this classification
simply corresponds to $w<4$ and $\overline w <4$,  so the observed cancellations follow directly from  \Eq{eq:gammaijzero}, as shown in Tab.~\ref{tab:dim6} .

\section{Weighing Tree Amplitudes}

We now compute the holomorphic and anti-holomorphic weights $(w_n,\overline w_n)$ for a general $n$-point on-shell tree amplitude in a renormalizable theory of massless particles.  We start with lower-point amplitudes and then apply induction to extend to higher-point.

Working in spinor helicity variables, we consider the three-point amplitude with coupling constant $g$,
\bea
A(1^{h_1}2^{h_2}3^{h_3}) = g \left\{
\begin{array}{ll}
\langle 12\rangle^{r_3}  \langle 23\rangle^{r_1} \langle 31\rangle^{r_2} ,& \, \sum_i h_i \leq0  \\
{[}12{]}^{\overline r_3}{[}23]^{\overline r_1}[31]^{\overline r_2}, &\,  \sum_i  h_i \geq0
 \end{array}
 \right.  \;\;\;  
 \label{eq:3pt}
\eea
which corresponds to MHV and $\overline{\textrm{MHV}}$ kinematics, $|1] \propto |2] \propto |3]$ and $|1\rangle \propto |2\rangle \propto |3\rangle$.
Lorentz invariance fixes the exponents to be
$ r_i = -\overline r_i=2h_i-\sum h$ and $\sum_i r_i = \sum_i\overline r_i=1-[g]$ by dimensional analysis~\cite{Benincasa:2007xk}.
According to \Eq{eq:3pt}, the weights of the three-point amplitude are
\bea
(w_3, \overline w_3) &=&  \left\{
\begin{array}{ll}
(4-[g],2+[g]),&\quad \sum_i h_i \leq0  \\
(2+[g],4-[g]), &\quad \sum_i  h_i \geq0 \end{array}
 \right. 
 \label{eq:3pt_w}
 \eea
In a renormalizable theory, $[g] =0$ or $1$, so we obtain
\bea
w_3, \overline w_3 \geq 2 ,
\eea
as a lower bound for the three-point amplitude.

Next, consider the four-point tree amplitude.  As we will see, $w_4, \overline w_4 \geq 4$ for the vast majority of amplitudes.  The reason is $w_4 < 4$ or $\overline w_4 <4$ requires non-zero total helicity, which is usually forbidden by helicity selection rules. To show this, we run through all possible candidate amplitudes with $w_4<4$. Analogous arguments of course apply for $\overline w_4 <4$.   

Most four-point tree amplitudes with $w_4=1$ or $3$  vanish because they have no Feynman diagrams, so
\begin{align}
0&= A(F^+ F^+ F^\pm \phi) =A(F^+ F^+ \psi^\pm \psi^\pm) \nn \\
&= A(F^+ F^- \psi^+ \psi^+)= A(F^+ \psi^+ \psi^- \phi)  \nn\\
&= A( \psi^+ \psi^+ \psi^+\psi^-). \nn
\end{align}
Furthermore, most amplitudes with $w_4 = 0$ or $2$ vanish due to helicity selection rules, so 
\begin{align}
0&= A(F^+ F^+ F^+ F^\pm)=A(F^+ F^+ \psi^+ \psi^-)\nn \\
&= A(F^+ F^+ \phi \; \phi) = A(F^+ \psi^+ \psi^+ \phi)  \nn.
\end{align}
While these amplitudes have Feynman diagrams, they vanish on-shell for their chosen helicities.
This leaves a handful of amplitudes that can in principle be non-zero,
\bea 
0 &\neq&  A(\psi^+ \psi^+ \psi^+ \psi^+), A(F^+ \phi \; \phi \; \phi), A(\psi^+ \psi^+ \phi \; \phi), \nn
\label{eq:except}
\eea
for which $w_4 = 2,3,3$, respectively.
These ``exceptional amplitudes'' are the only four-point tree amplitudes with $w_4 <4 $ that are not identically zero.

Since the exceptional amplitudes require external or internal scalars, they never arise in theories of only gauge bosons and fermions, {\it e.g.}~QCD.  The second and third amplitudes  require super-renormalizable cubic scalar interactions, which we do not consider here.  Meanwhile, the first amplitude arises from Yukawa couplings of non-holomorphic form, which is to say a combination of couplings of the form $\phi \psi^2$ together with $\bar \phi  \psi^2$.  In a supersymmetric theory, such couplings would violate holomorphy of the superpotential.  In the standard model, Higgs doublet exchange generates an exceptional amplitude proportional to the product up-type and down-type Yukawa couplings.  This diagram will be important later when we discuss renormalization in the standard model.    In summary, we find
\bea
w_4, \overline w_4 \geq 4,
\eea
as a lower bound for the four-point amplitude,  modulo the exceptional amplitudes.

  \begin{figure}[t]

\begin{center}
\includegraphics[width=.49\textwidth]{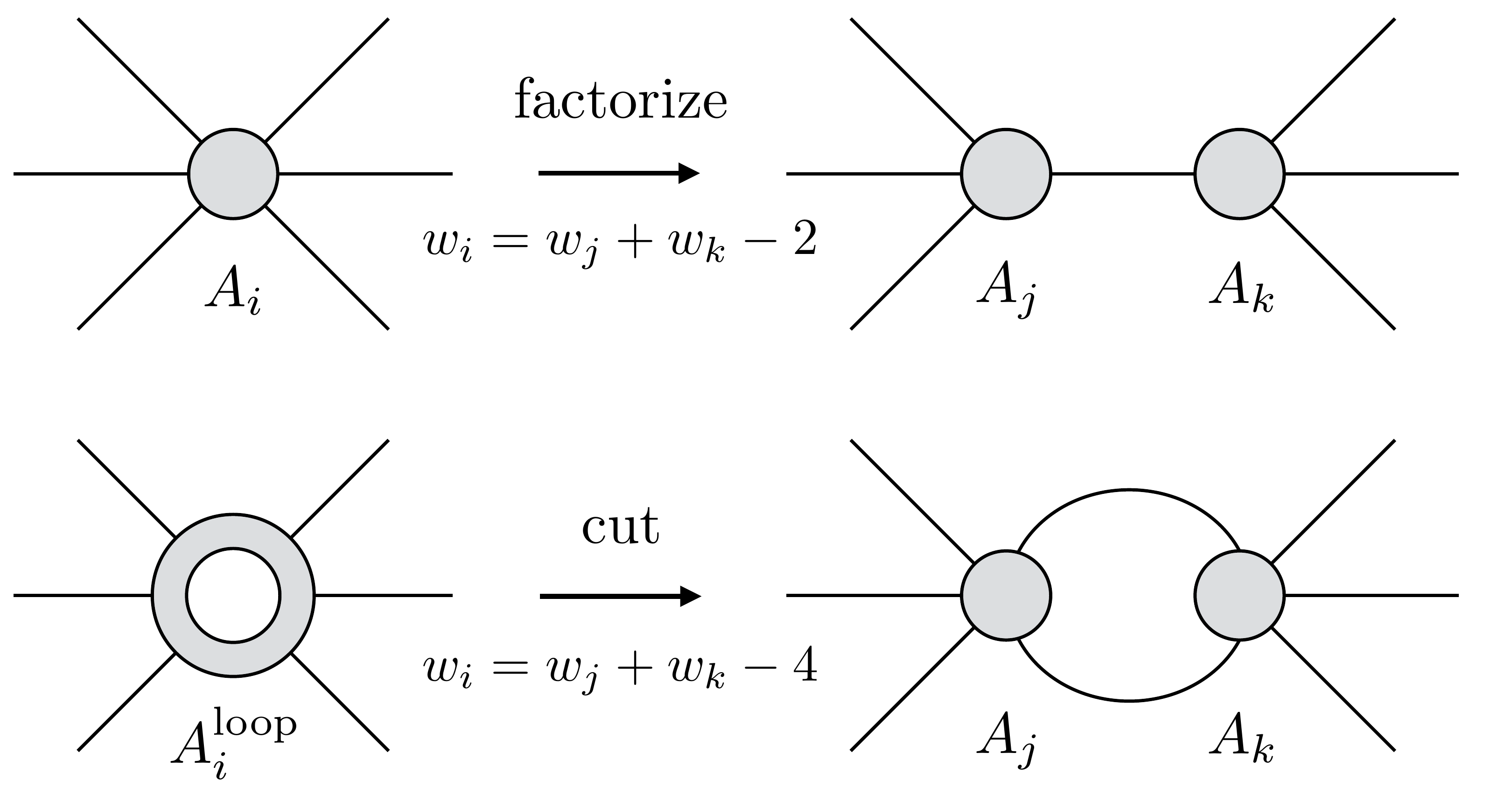}
\end{center}
\vspace*{-0.5cm}
\caption{Diagrams of tree factorization and one-loop unitarity, with the weight selection rules from \Eqs{eq:treerelation}{eq:looprelation}.}
\label{fig:amps} 
\end{figure}

Finally, consider a general higher-point tree amplitude, $A_i$, which on a factorization channel degenerates into a product of amplitudes, $A_j$ and $A_k$, where
\bea
{\rm fact}[A_i] &=&  \frac{i}{\ell^2}\sum_h A_j(\ell^h)A_k(-\ell^{-h}),
\eea
depicted in Fig.~\ref{fig:amps}.
If the total numbers and helicities of $A_i$, $A_j$, and $A_k$, are $(n_i,h_i)$, $(n_j,h_j)$, and $(n_k,h_k)$, then $
n_i = n_j + n_k -2$ and $h_i= h_j + h_k$ since each side of the factorization channel has an equal and opposite helicity.  Thus, the corresponding weights, 
$(w_i, \overline w_i)$, $(w_j, \overline w_j)$, and $(w_k, \overline w_k)$, satisfy the following tree selection rule,
\bea
\textrm{tree rule:}\qquad
\begin{array}{c}
w_i = w_{j} + w_{k} -2 \\
\overline w_i = \overline w_{j} + \overline w_{k} -2 
\end{array} 
\label{eq:treerelation}
\eea
We have already shown that $w_3,\overline w_3 \geq 2$ and $w_4, \overline w_4\geq 4$ modulo the exceptional diagrams.
Since all five-point amplitudes factorize into  three and four-point amplitudes, \Eq{eq:treerelation} implies that $w_5,\overline w_5 \geq 4$.  Induction to higher-point then yields the main result of this section,
\bea
w_n,\overline w_n &\geq& \left\{
\begin{array}{ll}
2, & \quad  n=3 \\
4, & \quad n > 3 \\
\end{array}
\right. 
\label{eq:weight_rule}
\eea
which, modulo exceptional amplitudes, is a lower bound on the weights of $n$-point tree amplitudes in a theory of massless particles with marginal interactions.  Note that even when exceptional amplitudes exist, $w_n,\overline{w}_n\geq 2$.

An important consequence of \Eq{eq:treerelation} is that attaching renormalizable interactions to any amplitude $A_j$, even involving irrelevant interactions, can only produce an amplitude $A_i$ of greater or equal weight.  To see why, note that $A_i$ factorizes into $A_j$ and an amplitude $A_k$ composed of only renormalizable interactions, where $w_k,\overline w_k \geq 2$ by \Eq{eq:weight_rule}. \Eq{eq:treerelation} then implies that $w_i \geq w_j$ and $\overline w_i \geq \overline w_j$, so the minimum weight amplitude involving a higher dimension operator is the contact amplitude built from a single insertion of that operator.


\section{Weighing One-Loop Amplitudes}

We now calculate the weights of one-loop amplitudes using generalized unitarity and the tree-level results of the previous section.  Leading order renormalization of higher dimension operators is characterized by the anomalous dimension matrix $\gamma_{ij}$, which encodes how ${\cal O}_i$ is radiatively generated by ${\cal O}_j$. In practice, $\gamma_{ij}$ is extracted from the one-loop amplitude $A^{\rm loop}_i$ involving an insertion of ${\cal O}_j$ that has precisely the same external states as the tree amplitude $A_i$ involving an insertion of ${\cal O}_i$. Any ultraviolet divergence in $A^{\rm loop}_i$ must be absorbed by the counterterm $A_i$.

The Passarino-Veltman (PV) reduction~\cite{Passarino:1978jh} of the one-loop amplitude $A^{\rm loop}_i$ is
\bea
A^{\rm loop}_i = \sum_{\rm box} d_4 I_4 +  \sum_{\rm triangle} d_3 I_3 +\sum_{\rm bubble} d_2 I_2  + \textrm{rational}, \nn
	\label{eq:PV_reduction}
\eea
summing over topologies of scalar box, triangle, and bubble integrals, $I_4$, $I_3$, and $I_2$.  Tadpole integrals vanish in massless limit considered here. The integral coefficients $d_4$, $d_3$, and $d_2$ are rational functions of external kinematic data.  Ultraviolet log divergences arise from the scalar bubble integrals in the PV reduction, where in dimensional regularization, $I_2 \rightarrow 1/ (4\pi)^2 \epsilon $.
Separating ultraviolet divergent and finite terms, we find
\bea
A^{\rm loop}_i  &=&\frac{1}{(4\pi)^2 \epsilon} \sum_{\rm bubble} d_2  + \textrm{finite} ,
\label{eq:divA}
\eea
which implies a counterterm tree amplitude,
\bea
A_i &=& -\frac{1}{(4\pi)^2 \epsilon} \sum_{\rm bubble} d_2 ,
\label{eq:count}
\eea
such that the sum, $A^{\rm loop}_i + A_i$, is finite.

With generalized unitarity~\cite{Bern:1994cg,*Bern:1994zx}, integral coefficients can be constructed by relating kinematic singularities of the one-loop amplitude to products of tree amplitudes.  In particular, the two-particle cut in a particular channel is
\bea
{\rm cut}[A^{\rm loop}_i] = \sum_{h_1,h_2} A_j(\ell_1^{h_1}, \ell_2^{h_2}) A_k(-\ell_1^{-h_1}, -\ell_2^{-h_2}),\quad
\label{eq:2cut1}
\eea
where $\ell_1, \ell_2$ and $h_1,h_2$ are the momenta and helicities of the cut lines and $A_j$ and $A_k$ are on-shell tree amplitudes corresponding to the cut channel, as  depicted in Fig.~\ref{fig:amps}.

Applying this same cut to the PV reduction, we find
\bea
{\rm cut}[A^{\rm loop}_i]  = d_2 + \textrm{terms that depend on } \ell_1,\ell_2 ,
\label{eq:2cut2}
\eea  
where the $\ell_1,\ell_2$ dependent terms come from two-particle cuts of triangle and box integrals.   As is well-known, the divergence of the one-loop amplitude is related to the two-particle cut \cite{ArkaniHamed:2008gz,Huang:2012aq,Dixon:2002}.  However, a kinematic singularity is present only if $A_j$ and $A_k$ are four-point amplitudes or higher, corresponding to ``massive'' bubble integrals. When $A_j$ or $A_k$ are three-point amplitudes, the associated ``massless'' bubble integrals are scaleless and vanish in dimensional regularization. For now we ignore these subtle contributions but revisit them in the next section.


Combining \Eq{eq:count} with \Eqs{eq:2cut1}{eq:2cut2}, we find that the total numbers and helicities $(n_i,h_i)$, $(n_j,h_j)$, $(n_k,h_k)$ of $A_i$, $A_j$ and $A_k$ satisfy
$n_i= n_j + n_k -4 $ and $h_i = h_j + h_k$.
This implies the one-loop selection rule,
\bea
\textrm{one-loop rule:}\qquad
\begin{array}{c}
w_i = w_{j} + w_{k} -4 \\
\overline w_i = \overline w_{j} + \overline w_{k} -4 
\end{array}
\label{eq:looprelation}
\eea
where $(w_i,\overline w_i)$, $(w_j, \overline w_j)$, and $(w_k, \overline w_k)$ are the weights of $A_i$, $A_j$, and $A_k$, respectively.  For the entry $\gamma_{ij}$ of the anomalous dimension matrix, we identify $A_i$ and $A_j$ with tree amplitudes built around insertions of ${\cal O}_i$ and ${\cal O}_j$, and $A_k$ with a tree amplitude of the renormalizable theory.    As noted earlier, the amplitudes on both sides of the cut must be four-point or higher to have a non-trivial unitarity cut, which implies from \Eq{eq:weight_rule} that $w_k,\overline w_k\geq 4$ if there are no exceptional amplitudes. Plugging back into \Eq{eq:looprelation} implies that $w_i \geq w_j$ and $\overline w_i \geq \overline w_j$, which is the non-renormalization theorem in \Eq{eq:nonrenorm}. If exceptional amplitudes are present, say from non-holomorphic Yukawas, then $w_k,\overline w_k=2$ and \Eq{eq:nonrenorm} is violated, albeit by exactly two units in weight.

\begin{table}[t]
	$\begin{array}{*2c| *3c| *3c| c|}
	& & F^2 \phi & F \psi^2 & \psi^2\phi^2 & \bar{F} \bar{\psi}^2 & \bar{F}^2 \phi &   \bar{\psi}^2\phi^2 & \phi^5  \\ 
	& (w,\bar{w})
	& (1,5) & (1,5) & (3,5) & (5,1) & (5,1) & (5,3) & (5,5) \\ \hline
	\rowcolor{zero2}
	\cw F^2 \phi & \cw (1,5)
	& \cw & \cw &  &  &  &  &  \\
	\rowcolor{zero2}
	\cw F \psi^2 & \cw (1,5)
	& \cw & \cw &  &  &  &  &  \\
	\rowcolor{zero2}
	\cw \psi^2\phi^2 & \cw (3,5)
	& \cw & \cw & \cw &  &  &  &  \\ \hline
	\rowcolor{zero2}
	\cw \bar{F}^2 \phi & \cw (5,1)
	&  &  &  & \cw & \cw &  &  \\
	\rowcolor{zero2}
	\cw \bar{F} \bar{\psi}^2 & \cw (5,1)
	&  &  &  & \cw & \cw &  &  \\
	\rowcolor{zero2}
	\cw \bar{\psi}^2\phi^2 & \cw (5,3)
	&  &  &  & \cw & \cw & \cw &  \\ \hline
	\rowcolor{white}
	\phi^5 & (5,5)
	&  &  &  &  &  &  &  \\ \hline
	\end{array}$
	\caption{Anomalous dimension matrix for dimension five operators in a general quantum field theory. The shaded entries vanish by our non-renormalization theorems. }
	\label{tab:dim5}
\end{table}

The weight lattice for all dimension five and six operators in a general quantum field theory is presented in \Fig{fig:wlattice}.  We use the operator basis of \cite{Grzadkowski:2010es} in which redundant operators, {\it e.g.}~those involving $\Box \phi$, are eliminated by equations of motion.   Our non-renormalization theorems imply that operators can only renormalize other operators of equal or greater weight, which in \Fig{fig:wlattice} forbids transitions that move down or to the left.  The form of the anomalous dimension matrix for all dimension five and six operators is shown in Tab.~\ref{tab:dim5} and Tab.~\ref{tab:dim6}.

\begin{table*}[t]
	$\begin{array} {*2c|  *5c| *5c| *4c| }
	& & F^{3} & F^{2} \phi^2 & F \psi^2 \phi & \psi^4 & 
	\psi^2 \phi^3 & 
	\bar{F}^3 & \bar{F}^2 \phi^2 & \bar{F}\bar{\psi}^{2} \phi & \bar{\psi}^4 & \bar{\psi}^{2} \phi^3 & \bar{\psi}^2 \psi^2 & \bar{\psi}\psi \phi^2 D & \phi^4 D^2 & \phi^6  \\
	& (w,\bar{w}) & (0,6) & (2,6) & (2,6) & (2,6) & (4,6) & (6,0) & (6,2) & (6,2) & (6,2) &
	(6,4) & (4,4) & (4,4) & (4,4) & (6,6) \\ \hline
	\rowcolor{zero2}
	\cw F^{3} & \cw (0,6) & \cw &  & \nda & \nda & \nda & \czb & \czb & \nda & \nda & \nda & \nda & \nda & \nda & \nda \\
	\rowcolor{zero2}
	\cw F^{2} \phi^2 & \cw (2,6) &\cw &\cw &\cw & \cw \nda & \nda & \czb & \czb & \czb & \nda & \nda & \nda & \czb & \czb & \nda \\
	\rowcolor{zero2}
	\cw F \psi^2 \phi & \cw (2,6) & \cw & \cw & \cw & \cw &  \czb & \czb & \czb & \czb & \nda & \czb & \czb & \czb & \nda & \nda \\
	\rowcolor{zero2}
	\cw \psi^4 & \cw (2,6) & \cw \nda & \cw \nda & \cw & \cw & \czb \nda & \czb \nda & \nda & \nda & \nda & \nda & \ysq & \czb & \nda & \nda  \\
	\rowcolor{white}
	\psi^2 \phi^3 & (4,6) & \cw {\nda^*} &  &  &  &  & \czb & \czb & \czb & \czb & \czb {\ysq} &  &  &  & \czb \nda \\ \hline
	\rowcolor{zero2}
	\cw\bar{F}^3 & \cw (6,0) & \czb & \czb & \nda & \nda & \nda & \cw & \czb & \nda & \nda & \nda & \nda & \nda & \nda & \nda \\
	\rowcolor{zero2}
	\cw\bar{F}^2 \phi^2 & \cw(6,2) & \czb & \czb & \czb & \nda & \nda & \cw & \cw & \cw & \cw \nda & \nda & \nda & \czb & \czb & \nda \\
	\rowcolor{zero2}
	\cw\bar{F}\bar{\psi}^{2} \phi & \cw(6,2) & \czb & \czb & \czb & \nda & \czb & \cw & \cw & \cw & \cw & \czb & \czb & \czb & \nda & \nda \\
	\rowcolor{zero2}
	\cw\bar{\psi}^4 & \cw(6,2) & \czb \nda & \nda & \nda & \nda & \nda & \cw \nda  & \cw \nda & \cw & \cw & \nda & \ydsq & \czb & \nda & \nda \\
	\rowcolor{white}
	\bar{\psi}^{2} \phi^3 & (6,4) & \czb & \czb & \czb & \czb & \czb{\ydsq } & \cw {\nda^*} &  &  &  &  &  &  &  & \czb \nda \\ \hline
	\rowcolor{zero2}
	\cw\bar{\psi}^2 \psi^2 & \cw(4,4) & \czb & \nda & \czb & \ydsq & \nda & \czb & \nda & \czb & \ysq & \nda & \cw & \cw & \cw \nda & \nda  \\
	\rowcolor{zero2}
	\cw\bar{\psi}\psi \phi^2 D & \cw(4,4) & \czb & \czb & \czb & \czb & \czb & \czb & \czb & \czb & \czb & \czb & \cw & \cw & \cw & \nda \\
	\rowcolor{zero2}
	\cw\phi^4 D^2 & \cw(4,4) & \czb & \czb & \czb & \nda & \czb & \czb & \czb & \czb & \nda & \czb & \cw \nda & \cw & \cw & \nda \\
	\rowcolor{white}
	\phi^6 & (6,6) & \cw {\nda^*} &  & \nda & \nda & & \cw {\nda^*} &  & \nda & \nda &  & \nda &  &  &  \\ \hline
	\end{array}$
\caption{Anomalous dimension matrix for dimension six operators in a general quantum field theory. The shaded entries vanish by our non-renormalization theorems, in full agreement with~\cite{Alonso:2014rga}. Here $y^2$ and $\bar y^2$ label entries that are non-zero due to non-holomorphic Yukawa couplings, $\times$ labels entries that vanish because there are no diagrams~\cite{Jenkins:2013sda}, and $\times^*$ labels entries that vanish by a combination of counterterm analysis and our non-renormalization theorems. }
\label{tab:dim6}
\end{table*}


\section{Infrared Divergences}

We now return to the issue of massless bubble integrals.  While these contributions formally vanish in dimensional regularization, this is potentially misleading because ultraviolet and infrared divergences enter with opposite sign $1/\epsilon$ poles.  Thus, an ultraviolet divergence may actually be present if there happens to be an equal and opposite virtual infrared divergence \cite{ArkaniHamed:2008gz,Huang:2012aq,Dixon:2002}.  Crucially, the Kinoshita-Lee-Nauenberg theorem~\cite{Kinoshita:1962ur,*Lee:1964is} says that all virtual infrared divergences are canceled by an inclusive sum over final states corresponding to tree-level real emission of an unresolved soft or collinear particle.  Inverting the logic, if real emission is actually infrared finite, then there can be no virtual infrared divergence and thus no ultraviolet divergence.  As we will show, this is true of the discarded contributions from massless bubbles which could have a priori violated \Eq{eq:nonrenorm}.

To diagnose potential infrared divergences in $A_i^{\rm loop}$, we analyze the associated amplitude for real emission, $A^{\rm real}_{i'}$.   In the infrared  regime, the singular part of this amplitude factorizes: $A^{\rm real}_{i'} \rightarrow A_i S_{i\rightarrow i'} + A_j S_{j\rightarrow i'}$, where $A_i$ and $A_j$ are tree amplitudes built around insertions of ${\cal O}_i$ and ${\cal O}_j$, and $S_{i\rightarrow i'}$ and $S_{j\rightarrow i'}$ are soft-collinear functions describing the emission of an unresolved particle.  Since infrared divergences are a long distance effect, we only consider the soft-collinear functions for emissions generated by marginal interactions.
They diverge as $1/\omega$ and $1/\sqrt{1-\cos \theta}$ in the soft and collinear limits, respectively, where $\omega$ and $\theta$ are the energy and splitting angle characterizing the emitted particle.   Since the phase-space measure is $ \int d\omega \,\omega \int d\cos\theta$, infrared divergences  require that $S_{i\rightarrow i'}$ and $S_{j\rightarrow i'}$ are both soft and/or collinear.

For soft emission, the hard process is unchanged \cite{Weinberg:1965nx}. Since $A_i S_{i\rightarrow i'}$ and $A_j S_{j\rightarrow i'}$ contribute to the same process, this implies that $A_i$ and $A_j$ have the same external states and thus equal weight, $w_i =w_j$.  While massless bubbles do contribute infrared and ultraviolet divergences not previously accounted for, this is perfectly consistent with the non-renormalization theorem in \Eq{eq:nonrenorm}, which allows for operator mixing when $w_i=w_j$.  Violation of \Eq{eq:nonrenorm} requires infrared divergences when $w_i < w_j$, but this only happens for soft emission that flips the helicity of a hard particle, which is subleading in the soft limit and thus finite upon $\int d \omega$ integration.

Similarly, collinear emission is divergent for $w_i=w_j$ but finite for $w_i < w_j$.  Since $A_i S_{i\rightarrow i'}$ and $A_j S_{j\rightarrow i'}$ have the same external states and weight, restricting to $w_i <w_j$ means that $w(S_{i\rightarrow i'})>w(S_{j\rightarrow i'})$.   \Eq{eq:3pt_w} then implies that $S_{i\rightarrow i'}$ and $S_{j\rightarrow i'}$  are collinear splitting functions generated by on-shell $\overline{\rm MHV}$ and MHV amplitudes.  As a result, the interference term  $S_{j\rightarrow i'}^* S_{i\rightarrow i'}$ carries net little group weight with respect to the mother particle initiating the collinear emission.  Rotations of angle $\phi$ around the axis of the mother particle act as a little group transformation on $S_{j\rightarrow i'}^* S_{i\rightarrow i'}$, yielding a net phase $e^{2i\phi}$ in the differential cross-section. Integrating over this angle yields $\int^{2\pi}_0 d\phi \, e^{2i\phi}=0$, so the collinear singularity vanishes upon phase-space integration. 

In summary, for $w_i < w_j$ we have found that real emission is infrared finite, so there are no ultraviolet divergences from massless bubbles.  The non-renormalization theorems in \Eq{eq:nonrenorm} apply despite infrared subtleties.

\section{Application to the Standard Model}

Since our results rely on unitarity and helicity, they apply to any four-dimensional quantum field theory of massless particles, including the standard model and its extension to higher dimension operators.
Incidentally, there has been much progress in this direction in recent years~\cite{Alonso:2014rga,Grojean:2013kd,Jenkins:2013zja,*Jenkins:2013wua,*Alonso:2013hga,Elias-Miro:2013gya,*Elias-Miro:2013mua,Elias-Miro:2013eta}.  A tour de force calculation of the full one-loop anomalous dimension matrix of dimension six operators  \cite{Jenkins:2013zja,*Jenkins:2013wua,*Alonso:2013hga} unearthed  a string of  miraculous cancellations not enforced by an obvious symmetry and visible only after the meticulous application of equations of motion \cite{Alonso:2014rga}.   Lacking a manifest symmetry of the Lagrangian, the authors of \cite{Alonso:2014rga} conjectured an underlying ``holomorphy'' of the standard model effective theory that ensures closure of certain operators under renormalization.  

The cancellations in \cite{Alonso:2014rga} are a direct consequence of the non-renormalization theorems in \Eq{eq:nonrenorm} and \Eq{eq:gammaijzero}, based on a classification of holomorphic ($w<4$), anti-holomorphic ($\overline w<4$), and non-holomorphic operators ($w,\overline w \geq 4$), and violated only by exceptional amplitudes ($w,\overline w =2$) generated by non-holomorphic Yukawas.  
The shaded entries in Tab.~\ref{tab:dim6} denote zeroes enforced by our non-renormalization theorems.  Entries marked with $\times$ vanish trivially because there are no associated Feynman diagrams, while the few entries marked with $\times^*$ vanish because the expected divergences in $\psi^2\phi^3$ and $\phi^6$ are accompanied by a counterterm of the form $\phi^4D^2$ \cite{Alonso:2013hga} which is forbidden by our non-renormalization theorems.

Interestingly,  the superfield formalism offers an enlightening albeit partial explanation of these cancellations \cite{Elias-Miro:2014eia} as well as analogous effects in chiral perturbation theory~\cite{Gasser:1983yg}.  These results are clearly connected to our own via the well-known ``effective'' supersymmetry of tree-level QCD~\cite{Parke:1985pn,*Kunszt:1985mg,*Dixon:1996wi,*Dixon:2010ik}, and so merits further study.



\section{Conclusions}

We have derived a new class of one-loop non-renormalization theorems for higher dimension operators in a general four-dimensional quantum field theory.  Since our arguments make no reference to symmetry---only unitarity and helicity---they are broadly applicable, and explain the peculiar cancellations observed in the renormalization of dimension six operators in the standard model.  Let us briefly discuss future directions.

First and foremost is the matter of higher loop orders.  As is well-known, helicity selection rules---{\it e.g.}~the vanishing of the all minus amplitude in Yang-Mills---are violated by finite one-loop corrections~\cite{Bern:1993mq,*Bern:1993qk,*Mahlon:1993si,*Bern:1994ju}.  While this strongly suggests that \Eq{eq:nonrenorm} should fail at two-loop order, this important question deserves close examination.  Another natural direction is higher dimensions, where helicity is naturally extended \cite{Cheung:2009dc} and dimensional reduction offers a bridge to massive theories.  Finally, there is the question of finding concrete linkage between our results and more conventional symmetry arguments like those of \cite{Elias-Miro:2014eia}.  Indeed, our definition of weight is reminiscent of both $R$-symmetry and twist, which are known to relate closely to existing non-renormalization theorems.  

 
\bigskip

\noindent {\it Acknowledgments}: 
We would like to thank Rodrigo Alonso, Zvi Bern, Lance Dixon, Yu-tin Huang, Elizabeth Jenkins, David Kosower, and Aneesh Manohar for useful discussions.  C.C. and C.-H.S. are supported by a DOE Early Career Award under Grant No.~DE-SC0010255. C.C. is also supported by a Sloan Research Fellowship.

\bibliographystyle{apsrev4-1}
 
\bibliography{holomorphyNotes}

\begin{thebibliography}{31}%
\makeatletter
\providecommand \@ifxundefined [1]{%
 \@ifx{#1\undefined}
}%
\providecommand \@ifnum [1]{%
 \ifnum #1\expandafter \@firstoftwo
 \else \expandafter \@secondoftwo
 \fi
}%
\providecommand \@ifx [1]{%
 \ifx #1\expandafter \@firstoftwo
 \else \expandafter \@secondoftwo
 \fi
}%
\providecommand \natexlab [1]{#1}%
\providecommand \enquote  [1]{``#1''}%
\providecommand \bibnamefont  [1]{#1}%
\providecommand \bibfnamefont [1]{#1}%
\providecommand \citenamefont [1]{#1}%
\providecommand \href@noop [0]{\@secondoftwo}%
\providecommand \href [0]{\begingroup \@sanitize@url \@href}%
\providecommand \@href[1]{\@@startlink{#1}\@@href}%
\providecommand \@@href[1]{\endgroup#1\@@endlink}%
\providecommand \@sanitize@url [0]{\catcode `\\12\catcode `\$12\catcode
  `\&12\catcode `\#12\catcode `\^12\catcode `\_12\catcode `\%12\relax}%
\providecommand \@@startlink[1]{}%
\providecommand \@@endlink[0]{}%
\providecommand \url  [0]{\begingroup\@sanitize@url \@url }%
\providecommand \@url [1]{\endgroup\@href {#1}{\urlprefix }}%
\providecommand \urlprefix  [0]{URL }%
\providecommand \Eprint [0]{\href }%
\providecommand \doibase [0]{http://dx.doi.org/}%
\providecommand \selectlanguage [0]{\@gobble}%
\providecommand \bibinfo  [0]{\@secondoftwo}%
\providecommand \bibfield  [0]{\@secondoftwo}%
\providecommand \translation [1]{[#1]}%
\providecommand \BibitemOpen [0]{}%
\providecommand \bibitemStop [0]{}%
\providecommand \bibitemNoStop [0]{.\EOS\space}%
\providecommand \EOS [0]{\spacefactor3000\relax}%
\providecommand \BibitemShut  [1]{\csname bibitem#1\endcsname}%
\let\auto@bib@innerbib\@empty
\bibitem [{\citenamefont {Bern}\ \emph {et~al.}(1995)\citenamefont {Bern},
  \citenamefont {Dixon}, \citenamefont {Dunbar},\ and\ \citenamefont
  {Kosower}}]{Bern:1994cg}%
  \BibitemOpen
  \bibfield  {author} {\bibinfo {author} {\bibfnamefont {Z.}~\bibnamefont
  {Bern}}, \bibinfo {author} {\bibfnamefont {L.~J.}\ \bibnamefont {Dixon}},
  \bibinfo {author} {\bibfnamefont {D.~C.}\ \bibnamefont {Dunbar}}, \ and\
  \bibinfo {author} {\bibfnamefont {D.~A.}\ \bibnamefont {Kosower}},\ }\href
  {\doibase 10.1016/0550-3213(94)00488-Z} {\bibfield  {journal} {\bibinfo
  {journal} {Nucl.Phys.}\ }\textbf {\bibinfo {volume} {B435}},\ \bibinfo
  {pages} {59} (\bibinfo {year} {1995})},\ \Eprint
  {http://arxiv.org/abs/hep-ph/9409265} {arXiv:hep-ph/9409265 [hep-ph]}
  \BibitemShut {NoStop}%
\bibitem [{\citenamefont {Bern}\ \emph
  {et~al.}(1994{\natexlab{a}})\citenamefont {Bern}, \citenamefont {Dixon},
  \citenamefont {Dunbar},\ and\ \citenamefont {Kosower}}]{Bern:1994zx}%
  \BibitemOpen
  \bibfield  {author} {\bibinfo {author} {\bibfnamefont {Z.}~\bibnamefont
  {Bern}}, \bibinfo {author} {\bibfnamefont {L.~J.}\ \bibnamefont {Dixon}},
  \bibinfo {author} {\bibfnamefont {D.~C.}\ \bibnamefont {Dunbar}}, \ and\
  \bibinfo {author} {\bibfnamefont {D.~A.}\ \bibnamefont {Kosower}},\ }\href
  {\doibase 10.1016/0550-3213(94)90179-1} {\bibfield  {journal} {\bibinfo
  {journal} {Nucl.Phys.}\ }\textbf {\bibinfo {volume} {B425}},\ \bibinfo
  {pages} {217} (\bibinfo {year} {1994}{\natexlab{a}})},\ \Eprint
  {http://arxiv.org/abs/hep-ph/9403226} {arXiv:hep-ph/9403226 [hep-ph]}
  \BibitemShut {NoStop}%
\bibitem [{\citenamefont {Alonso}\ \emph
  {et~al.}(2014{\natexlab{a}})\citenamefont {Alonso}, \citenamefont {Jenkins},\
  and\ \citenamefont {Manohar}}]{Alonso:2014rga}%
  \BibitemOpen
  \bibfield  {author} {\bibinfo {author} {\bibfnamefont {R.}~\bibnamefont
  {Alonso}}, \bibinfo {author} {\bibfnamefont {E.~E.}\ \bibnamefont {Jenkins}},
  \ and\ \bibinfo {author} {\bibfnamefont {A.~V.}\ \bibnamefont {Manohar}},\
  }\href {\doibase 10.1016/j.physletb.2014.10.045} {\bibfield  {journal}
  {\bibinfo  {journal} {Phys.Lett.}\ }\textbf {\bibinfo {volume} {B739}},\
  \bibinfo {pages} {95} (\bibinfo {year} {2014}{\natexlab{a}})},\ \Eprint
  {http://arxiv.org/abs/1409.0868} {arXiv:1409.0868 [hep-ph]} \BibitemShut
  {NoStop}%
\bibitem [{\citenamefont {Grojean}\ \emph {et~al.}(2013)\citenamefont
  {Grojean}, \citenamefont {Jenkins}, \citenamefont {Manohar},\ and\
  \citenamefont {Trott}}]{Grojean:2013kd}%
  \BibitemOpen
  \bibfield  {author} {\bibinfo {author} {\bibfnamefont {C.}~\bibnamefont
  {Grojean}}, \bibinfo {author} {\bibfnamefont {E.~E.}\ \bibnamefont
  {Jenkins}}, \bibinfo {author} {\bibfnamefont {A.~V.}\ \bibnamefont
  {Manohar}}, \ and\ \bibinfo {author} {\bibfnamefont {M.}~\bibnamefont
  {Trott}},\ }\href {\doibase 10.1007/JHEP04(2013)016} {\bibfield  {journal}
  {\bibinfo  {journal} {JHEP}\ }\textbf {\bibinfo {volume} {1304}},\ \bibinfo
  {pages} {016} (\bibinfo {year} {2013})},\ \Eprint
  {http://arxiv.org/abs/1301.2588} {arXiv:1301.2588 [hep-ph]} \BibitemShut
  {NoStop}%
\bibitem [{\citenamefont {Jenkins}\ \emph
  {et~al.}(2013{\natexlab{a}})\citenamefont {Jenkins}, \citenamefont
  {Manohar},\ and\ \citenamefont {Trott}}]{Jenkins:2013zja}%
  \BibitemOpen
  \bibfield  {author} {\bibinfo {author} {\bibfnamefont {E.~E.}\ \bibnamefont
  {Jenkins}}, \bibinfo {author} {\bibfnamefont {A.~V.}\ \bibnamefont
  {Manohar}}, \ and\ \bibinfo {author} {\bibfnamefont {M.}~\bibnamefont
  {Trott}},\ }\href {\doibase 10.1007/JHEP10(2013)087} {\bibfield  {journal}
  {\bibinfo  {journal} {JHEP}\ }\textbf {\bibinfo {volume} {1310}},\ \bibinfo
  {pages} {087} (\bibinfo {year} {2013}{\natexlab{a}})},\ \Eprint
  {http://arxiv.org/abs/1308.2627} {arXiv:1308.2627 [hep-ph]} \BibitemShut
  {NoStop}%
\bibitem [{\citenamefont {Jenkins}\ \emph {et~al.}(2014)\citenamefont
  {Jenkins}, \citenamefont {Manohar},\ and\ \citenamefont
  {Trott}}]{Jenkins:2013wua}%
  \BibitemOpen
  \bibfield  {author} {\bibinfo {author} {\bibfnamefont {E.~E.}\ \bibnamefont
  {Jenkins}}, \bibinfo {author} {\bibfnamefont {A.~V.}\ \bibnamefont
  {Manohar}}, \ and\ \bibinfo {author} {\bibfnamefont {M.}~\bibnamefont
  {Trott}},\ }\href {\doibase 10.1007/JHEP01(2014)035} {\bibfield  {journal}
  {\bibinfo  {journal} {JHEP}\ }\textbf {\bibinfo {volume} {1401}},\ \bibinfo
  {pages} {035} (\bibinfo {year} {2014})},\ \Eprint
  {http://arxiv.org/abs/1310.4838} {arXiv:1310.4838 [hep-ph]} \BibitemShut
  {NoStop}%
\bibitem [{\citenamefont {Alonso}\ \emph
  {et~al.}(2014{\natexlab{b}})\citenamefont {Alonso}, \citenamefont {Jenkins},
  \citenamefont {Manohar},\ and\ \citenamefont {Trott}}]{Alonso:2013hga}%
  \BibitemOpen
  \bibfield  {author} {\bibinfo {author} {\bibfnamefont {R.}~\bibnamefont
  {Alonso}}, \bibinfo {author} {\bibfnamefont {E.~E.}\ \bibnamefont {Jenkins}},
  \bibinfo {author} {\bibfnamefont {A.~V.}\ \bibnamefont {Manohar}}, \ and\
  \bibinfo {author} {\bibfnamefont {M.}~\bibnamefont {Trott}},\ }\href
  {\doibase 10.1007/JHEP04(2014)159} {\bibfield  {journal} {\bibinfo  {journal}
  {JHEP}\ }\textbf {\bibinfo {volume} {1404}},\ \bibinfo {pages} {159}
  (\bibinfo {year} {2014}{\natexlab{b}})},\ \Eprint
  {http://arxiv.org/abs/1312.2014} {arXiv:1312.2014 [hep-ph]} \BibitemShut
  {NoStop}%
\bibitem [{\citenamefont {Elias-Mir\'o}\ \emph
  {et~al.}(2013{\natexlab{a}})\citenamefont {Elias-Mir\'o}, \citenamefont
  {Espinosa}, \citenamefont {Masso},\ and\ \citenamefont
  {Pomarol}}]{Elias-Miro:2013gya}%
  \BibitemOpen
  \bibfield  {author} {\bibinfo {author} {\bibfnamefont {J.}~\bibnamefont
  {Elias-Mir\'o}}, \bibinfo {author} {\bibfnamefont {J.}~\bibnamefont
  {Espinosa}}, \bibinfo {author} {\bibfnamefont {E.}~\bibnamefont {Masso}}, \
  and\ \bibinfo {author} {\bibfnamefont {A.}~\bibnamefont {Pomarol}},\ }\href
  {\doibase 10.1007/JHEP08(2013)033} {\bibfield  {journal} {\bibinfo  {journal}
  {JHEP}\ }\textbf {\bibinfo {volume} {1308}},\ \bibinfo {pages} {033}
  (\bibinfo {year} {2013}{\natexlab{a}})},\ \Eprint
  {http://arxiv.org/abs/1302.5661} {arXiv:1302.5661 [hep-ph]} \BibitemShut
  {NoStop}%
\bibitem [{\citenamefont {Elias-Mir\'o}\ \emph
  {et~al.}(2013{\natexlab{b}})\citenamefont {Elias-Mir\'o}, \citenamefont
  {Espinosa}, \citenamefont {Masso},\ and\ \citenamefont
  {Pomarol}}]{Elias-Miro:2013mua}%
  \BibitemOpen
  \bibfield  {author} {\bibinfo {author} {\bibfnamefont {J.}~\bibnamefont
  {Elias-Mir\'o}}, \bibinfo {author} {\bibfnamefont {J.}~\bibnamefont
  {Espinosa}}, \bibinfo {author} {\bibfnamefont {E.}~\bibnamefont {Masso}}, \
  and\ \bibinfo {author} {\bibfnamefont {A.}~\bibnamefont {Pomarol}},\ }\href
  {\doibase 10.1007/JHEP11(2013)066} {\bibfield  {journal} {\bibinfo  {journal}
  {JHEP}\ }\textbf {\bibinfo {volume} {1311}},\ \bibinfo {pages} {066}
  (\bibinfo {year} {2013}{\natexlab{b}})},\ \Eprint
  {http://arxiv.org/abs/1308.1879} {arXiv:1308.1879 [hep-ph]} \BibitemShut
  {NoStop}%
\bibitem [{\citenamefont {Elias-Mir\'o}\ \emph
  {et~al.}(2014{\natexlab{a}})\citenamefont {Elias-Mir\'o}, \citenamefont
  {Grojean}, \citenamefont {Gupta},\ and\ \citenamefont
  {Marzocca}}]{Elias-Miro:2013eta}%
  \BibitemOpen
  \bibfield  {author} {\bibinfo {author} {\bibfnamefont {J.}~\bibnamefont
  {Elias-Mir\'o}}, \bibinfo {author} {\bibfnamefont {C.}~\bibnamefont
  {Grojean}}, \bibinfo {author} {\bibfnamefont {R.~S.}\ \bibnamefont {Gupta}},
  \ and\ \bibinfo {author} {\bibfnamefont {D.}~\bibnamefont {Marzocca}},\
  }\href {\doibase 10.1007/JHEP05(2014)019} {\bibfield  {journal} {\bibinfo
  {journal} {JHEP}\ }\textbf {\bibinfo {volume} {1405}},\ \bibinfo {pages}
  {019} (\bibinfo {year} {2014}{\natexlab{a}})},\ \Eprint
  {http://arxiv.org/abs/1312.2928} {arXiv:1312.2928 [hep-ph]} \BibitemShut
  {NoStop}%
\bibitem [{\citenamefont {Benincasa}\ and\ \citenamefont
  {Cachazo}(2007)}]{Benincasa:2007xk}%
  \BibitemOpen
  \bibfield  {author} {\bibinfo {author} {\bibfnamefont {P.}~\bibnamefont
  {Benincasa}}\ and\ \bibinfo {author} {\bibfnamefont {F.}~\bibnamefont
  {Cachazo}},\ }\href@noop {} {\  (\bibinfo {year} {2007})},\ \Eprint
  {http://arxiv.org/abs/0705.4305} {arXiv:0705.4305 [hep-th]} \BibitemShut
  {NoStop}%
\bibitem [{\citenamefont {Passarino}\ and\ \citenamefont
  {Veltman}(1979)}]{Passarino:1978jh}%
  \BibitemOpen
  \bibfield  {author} {\bibinfo {author} {\bibfnamefont {G.}~\bibnamefont
  {Passarino}}\ and\ \bibinfo {author} {\bibfnamefont {M.}~\bibnamefont
  {Veltman}},\ }\href {\doibase 10.1016/0550-3213(79)90234-7} {\bibfield
  {journal} {\bibinfo  {journal} {Nucl.Phys.}\ }\textbf {\bibinfo {volume}
  {B160}},\ \bibinfo {pages} {151} (\bibinfo {year} {1979})}\BibitemShut
  {NoStop}%
\bibitem [{\citenamefont {Arkani-Hamed}\ \emph {et~al.}(2010)\citenamefont
  {Arkani-Hamed}, \citenamefont {Cachazo},\ and\ \citenamefont
  {Kaplan}}]{ArkaniHamed:2008gz}%
  \BibitemOpen
  \bibfield  {author} {\bibinfo {author} {\bibfnamefont {N.}~\bibnamefont
  {Arkani-Hamed}}, \bibinfo {author} {\bibfnamefont {F.}~\bibnamefont
  {Cachazo}}, \ and\ \bibinfo {author} {\bibfnamefont {J.}~\bibnamefont
  {Kaplan}},\ }\href {\doibase 10.1007/JHEP09(2010)016} {\bibfield  {journal}
  {\bibinfo  {journal} {JHEP}\ }\textbf {\bibinfo {volume} {1009}},\ \bibinfo
  {pages} {016} (\bibinfo {year} {2010})},\ \Eprint
  {http://arxiv.org/abs/0808.1446} {arXiv:0808.1446 [hep-th]} \BibitemShut
  {NoStop}%
\bibitem [{\citenamefont {Huang}\ \emph {et~al.}(2013)\citenamefont {Huang},
  \citenamefont {McGady},\ and\ \citenamefont {Peng}}]{Huang:2012aq}%
  \BibitemOpen
  \bibfield  {author} {\bibinfo {author} {\bibfnamefont {Y.-t.}\ \bibnamefont
  {Huang}}, \bibinfo {author} {\bibfnamefont {D.~A.}\ \bibnamefont {McGady}}, \
  and\ \bibinfo {author} {\bibfnamefont {C.}~\bibnamefont {Peng}},\ }\href
  {\doibase 10.1103/PhysRevD.87.085028} {\bibfield  {journal} {\bibinfo
  {journal} {Phys.Rev.}\ }\textbf {\bibinfo {volume} {D87}},\ \bibinfo {pages}
  {085028} (\bibinfo {year} {2013})},\ \Eprint {http://arxiv.org/abs/1205.5606}
  {arXiv:1205.5606 [hep-th]} \BibitemShut {NoStop}%
\bibitem [{\citenamefont {Dixon}()}]{Dixon:2002}%
  \BibitemOpen
  \bibfield  {author} {\bibinfo {author} {\bibfnamefont {L.}~\bibnamefont
  {Dixon}},\ }\href@noop {} {\bibinfo  {journal} {private communication}\
  }\BibitemShut {NoStop}%
\bibitem [{\citenamefont {Grzadkowski}\ \emph {et~al.}(2010)\citenamefont
  {Grzadkowski}, \citenamefont {Iskrzynski}, \citenamefont {Misiak},\ and\
  \citenamefont {Rosiek}}]{Grzadkowski:2010es}%
  \BibitemOpen
\bibfield  {journal} {  }\bibfield  {author} {\bibinfo {author} {\bibfnamefont
  {B.}~\bibnamefont {Grzadkowski}}, \bibinfo {author} {\bibfnamefont
  {M.}~\bibnamefont {Iskrzynski}}, \bibinfo {author} {\bibfnamefont
  {M.}~\bibnamefont {Misiak}}, \ and\ \bibinfo {author} {\bibfnamefont
  {J.}~\bibnamefont {Rosiek}},\ }\href {\doibase 10.1007/JHEP10(2010)085}
  {\bibfield  {journal} {\bibinfo  {journal} {JHEP}\ }\textbf {\bibinfo
  {volume} {1010}},\ \bibinfo {pages} {085} (\bibinfo {year} {2010})},\ \Eprint
  {http://arxiv.org/abs/1008.4884} {arXiv:1008.4884 [hep-ph]} \BibitemShut
  {NoStop}%
\bibitem [{\citenamefont {Jenkins}\ \emph
  {et~al.}(2013{\natexlab{b}})\citenamefont {Jenkins}, \citenamefont
  {Manohar},\ and\ \citenamefont {Trott}}]{Jenkins:2013sda}%
  \BibitemOpen
  \bibfield  {author} {\bibinfo {author} {\bibfnamefont {E.~E.}\ \bibnamefont
  {Jenkins}}, \bibinfo {author} {\bibfnamefont {A.~V.}\ \bibnamefont
  {Manohar}}, \ and\ \bibinfo {author} {\bibfnamefont {M.}~\bibnamefont
  {Trott}},\ }\href {\doibase 10.1016/j.physletb.2013.09.020} {\bibfield
  {journal} {\bibinfo  {journal} {Phys.Lett.}\ }\textbf {\bibinfo {volume}
  {B726}},\ \bibinfo {pages} {697} (\bibinfo {year} {2013}{\natexlab{b}})},\
  \Eprint {http://arxiv.org/abs/1309.0819} {arXiv:1309.0819 [hep-ph]}
  \BibitemShut {NoStop}%
\bibitem [{\citenamefont {Kinoshita}(1962)}]{Kinoshita:1962ur}%
  \BibitemOpen
  \bibfield  {author} {\bibinfo {author} {\bibfnamefont {T.}~\bibnamefont
  {Kinoshita}},\ }\href {\doibase 10.1063/1.1724268} {\bibfield  {journal}
  {\bibinfo  {journal} {J.Math.Phys.}\ }\textbf {\bibinfo {volume} {3}},\
  \bibinfo {pages} {650} (\bibinfo {year} {1962})}\BibitemShut {NoStop}%
\bibitem [{\citenamefont {Lee}\ and\ \citenamefont
  {Nauenberg}(1964)}]{Lee:1964is}%
  \BibitemOpen
  \bibfield  {author} {\bibinfo {author} {\bibfnamefont {T.}~\bibnamefont
  {Lee}}\ and\ \bibinfo {author} {\bibfnamefont {M.}~\bibnamefont
  {Nauenberg}},\ }\href {\doibase 10.1103/PhysRev.133.B1549} {\bibfield
  {journal} {\bibinfo  {journal} {Phys.Rev.}\ }\textbf {\bibinfo {volume}
  {133}},\ \bibinfo {pages} {B1549} (\bibinfo {year} {1964})}\BibitemShut
  {NoStop}%
\bibitem [{\citenamefont {Weinberg}(1965)}]{Weinberg:1965nx}%
  \BibitemOpen
  \bibfield  {author} {\bibinfo {author} {\bibfnamefont {S.}~\bibnamefont
  {Weinberg}},\ }\href {\doibase 10.1103/PhysRev.140.B516} {\bibfield
  {journal} {\bibinfo  {journal} {Phys.Rev.}\ }\textbf {\bibinfo {volume}
  {140}},\ \bibinfo {pages} {B516} (\bibinfo {year} {1965})}\BibitemShut
  {NoStop}%
\bibitem [{\citenamefont {Elias-Mir\'o}\ \emph
  {et~al.}(2014{\natexlab{b}})\citenamefont {Elias-Mir\'o}, \citenamefont
  {Espinosa},\ and\ \citenamefont {Pomarol}}]{Elias-Miro:2014eia}%
  \BibitemOpen
  \bibfield  {author} {\bibinfo {author} {\bibfnamefont {J.}~\bibnamefont
  {Elias-Mir\'o}}, \bibinfo {author} {\bibfnamefont {J.}~\bibnamefont
  {Espinosa}}, \ and\ \bibinfo {author} {\bibfnamefont {A.}~\bibnamefont
  {Pomarol}},\ }\href@noop {} {\  (\bibinfo {year} {2014}{\natexlab{b}})},\
  \Eprint {http://arxiv.org/abs/1412.7151} {arXiv:1412.7151 [hep-ph]}
  \BibitemShut {NoStop}%
\bibitem [{\citenamefont {Gasser}\ and\ \citenamefont
  {Leutwyler}(1984)}]{Gasser:1983yg}%
  \BibitemOpen
  \bibfield  {author} {\bibinfo {author} {\bibfnamefont {J.}~\bibnamefont
  {Gasser}}\ and\ \bibinfo {author} {\bibfnamefont {H.}~\bibnamefont
  {Leutwyler}},\ }\href {\doibase 10.1016/0003-4916(84)90242-2} {\bibfield
  {journal} {\bibinfo  {journal} {Annals Phys.}\ }\textbf {\bibinfo {volume}
  {158}},\ \bibinfo {pages} {142} (\bibinfo {year} {1984})}\BibitemShut
  {NoStop}%
\bibitem [{\citenamefont {Parke}\ and\ \citenamefont
  {Taylor}(1985)}]{Parke:1985pn}%
  \BibitemOpen
  \bibfield  {author} {\bibinfo {author} {\bibfnamefont {S.~J.}\ \bibnamefont
  {Parke}}\ and\ \bibinfo {author} {\bibfnamefont {T.}~\bibnamefont {Taylor}},\
  }\href {\doibase 10.1016/0370-2693(85)91216-X} {\bibfield  {journal}
  {\bibinfo  {journal} {Phys.Lett.}\ }\textbf {\bibinfo {volume} {B157}},\
  \bibinfo {pages} {81} (\bibinfo {year} {1985})}\BibitemShut {NoStop}%
\bibitem [{\citenamefont {Kunszt}(1986)}]{Kunszt:1985mg}%
  \BibitemOpen
  \bibfield  {author} {\bibinfo {author} {\bibfnamefont {Z.}~\bibnamefont
  {Kunszt}},\ }\href {\doibase 10.1016/0550-3213(86)90319-6} {\bibfield
  {journal} {\bibinfo  {journal} {Nucl.Phys.}\ }\textbf {\bibinfo {volume}
  {B271}},\ \bibinfo {pages} {333} (\bibinfo {year} {1986})}\BibitemShut
  {NoStop}%
\bibitem [{\citenamefont {Dixon}(1996)}]{Dixon:1996wi}%
  \BibitemOpen
  \bibfield  {author} {\bibinfo {author} {\bibfnamefont {L.~J.}\ \bibnamefont
  {Dixon}},\ }\href@noop {} {\  (\bibinfo {year} {1996})},\ \Eprint
  {http://arxiv.org/abs/hep-ph/9601359} {arXiv:hep-ph/9601359 [hep-ph]}
  \BibitemShut {NoStop}%
\bibitem [{\citenamefont {Dixon}\ \emph {et~al.}(2011)\citenamefont {Dixon},
  \citenamefont {Henn}, \citenamefont {Plefka},\ and\ \citenamefont
  {Schuster}}]{Dixon:2010ik}%
  \BibitemOpen
  \bibfield  {author} {\bibinfo {author} {\bibfnamefont {L.~J.}\ \bibnamefont
  {Dixon}}, \bibinfo {author} {\bibfnamefont {J.~M.}\ \bibnamefont {Henn}},
  \bibinfo {author} {\bibfnamefont {J.}~\bibnamefont {Plefka}}, \ and\ \bibinfo
  {author} {\bibfnamefont {T.}~\bibnamefont {Schuster}},\ }\href {\doibase
  10.1007/JHEP01(2011)035} {\bibfield  {journal} {\bibinfo  {journal} {JHEP}\
  }\textbf {\bibinfo {volume} {1101}},\ \bibinfo {pages} {035} (\bibinfo {year}
  {2011})},\ \Eprint {http://arxiv.org/abs/1010.3991} {arXiv:1010.3991
  [hep-ph]} \BibitemShut {NoStop}%
\bibitem [{\citenamefont {Bern}\ \emph {et~al.}(1993)\citenamefont {Bern},
  \citenamefont {Dixon},\ and\ \citenamefont {Kosower}}]{Bern:1993mq}%
  \BibitemOpen
  \bibfield  {author} {\bibinfo {author} {\bibfnamefont {Z.}~\bibnamefont
  {Bern}}, \bibinfo {author} {\bibfnamefont {L.~J.}\ \bibnamefont {Dixon}}, \
  and\ \bibinfo {author} {\bibfnamefont {D.~A.}\ \bibnamefont {Kosower}},\
  }\href {\doibase 10.1103/PhysRevLett.70.2677} {\bibfield  {journal} {\bibinfo
   {journal} {Phys.Rev.Lett.}\ }\textbf {\bibinfo {volume} {70}},\ \bibinfo
  {pages} {2677} (\bibinfo {year} {1993})},\ \Eprint
  {http://arxiv.org/abs/hep-ph/9302280} {arXiv:hep-ph/9302280 [hep-ph]}
  \BibitemShut {NoStop}%
\bibitem [{\citenamefont {Bern}\ \emph
  {et~al.}(1994{\natexlab{b}})\citenamefont {Bern}, \citenamefont {Chalmers},
  \citenamefont {Dixon},\ and\ \citenamefont {Kosower}}]{Bern:1993qk}%
  \BibitemOpen
  \bibfield  {author} {\bibinfo {author} {\bibfnamefont {Z.}~\bibnamefont
  {Bern}}, \bibinfo {author} {\bibfnamefont {G.}~\bibnamefont {Chalmers}},
  \bibinfo {author} {\bibfnamefont {L.~J.}\ \bibnamefont {Dixon}}, \ and\
  \bibinfo {author} {\bibfnamefont {D.~A.}\ \bibnamefont {Kosower}},\ }\href
  {\doibase 10.1103/PhysRevLett.72.2134} {\bibfield  {journal} {\bibinfo
  {journal} {Phys.Rev.Lett.}\ }\textbf {\bibinfo {volume} {72}},\ \bibinfo
  {pages} {2134} (\bibinfo {year} {1994}{\natexlab{b}})},\ \Eprint
  {http://arxiv.org/abs/hep-ph/9312333} {arXiv:hep-ph/9312333 [hep-ph]}
  \BibitemShut {NoStop}%
\bibitem [{\citenamefont {Mahlon}(1994)}]{Mahlon:1993si}%
  \BibitemOpen
  \bibfield  {author} {\bibinfo {author} {\bibfnamefont {G.}~\bibnamefont
  {Mahlon}},\ }\href {\doibase 10.1103/PhysRevD.49.4438} {\bibfield  {journal}
  {\bibinfo  {journal} {Phys.Rev.}\ }\textbf {\bibinfo {volume} {D49}},\
  \bibinfo {pages} {4438} (\bibinfo {year} {1994})},\ \Eprint
  {http://arxiv.org/abs/hep-ph/9312276} {arXiv:hep-ph/9312276 [hep-ph]}
  \BibitemShut {NoStop}%
\bibitem [{\citenamefont {Bern}\ \emph
  {et~al.}(1994{\natexlab{c}})\citenamefont {Bern}, \citenamefont {Dixon},
  \citenamefont {Dunbar},\ and\ \citenamefont {Kosower}}]{Bern:1994ju}%
  \BibitemOpen
  \bibfield  {author} {\bibinfo {author} {\bibfnamefont {Z.}~\bibnamefont
  {Bern}}, \bibinfo {author} {\bibfnamefont {L.~J.}\ \bibnamefont {Dixon}},
  \bibinfo {author} {\bibfnamefont {D.~C.}\ \bibnamefont {Dunbar}}, \ and\
  \bibinfo {author} {\bibfnamefont {D.~A.}\ \bibnamefont {Kosower}},\
  }\href@noop {} {\  (\bibinfo {year} {1994}{\natexlab{c}})},\ \Eprint
  {http://arxiv.org/abs/hep-ph/9405248} {arXiv:hep-ph/9405248 [hep-ph]}
  \BibitemShut {NoStop}%
\bibitem [{\citenamefont {Cheung}\ and\ \citenamefont
  {O'Connell}(2009)}]{Cheung:2009dc}%
  \BibitemOpen
  \bibfield  {author} {\bibinfo {author} {\bibfnamefont {C.}~\bibnamefont
  {Cheung}}\ and\ \bibinfo {author} {\bibfnamefont {D.}~\bibnamefont
  {O'Connell}},\ }\href {\doibase 10.1088/1126-6708/2009/07/075} {\bibfield
  {journal} {\bibinfo  {journal} {JHEP}\ }\textbf {\bibinfo {volume} {0907}},\
  \bibinfo {pages} {075} (\bibinfo {year} {2009})},\ \Eprint
  {http://arxiv.org/abs/0902.0981} {arXiv:0902.0981 [hep-th]} \BibitemShut
  {NoStop}%
\end{thebibliography}%

\end{document}